\documentclass[12pt]{iopart}
\usepackage{epsf}

\begin{document}

\title[]{Test of Chemical freeze-out at RHIC}

\author{J. Takahashi {\it for the STAR Collaboration}}
\address{Instituto de F\'{i}sica Gleb Wataghin, University of Campinas - UNICAMP, Brazil}
\ead{jun@ifi.unicamp.br}

\begin{abstract}
We present the results of a systematic test applying statistical thermal model
fits in a consistent way for different particle ratios, and different
system sizes using the various particle yields measured in the STAR
experiment.
Comparison between central and peripheral Au+Au and Cu+Cu collisions
with data from p+p collisions provides an interesting tool to verify the
dependence with the system size. We also present a study of the rapidity 
dependence of the thermal fit parameters using available data from RHIC 
in the forward rapidity regions and also using different parameterization 
for the rapidity distribution of different particles.
\end{abstract}

\section{Introduction}

Statistical thermal model has been very successful in describing the various particle ratios 
measured in heavy ion collisions, from the SPS energies up to the highest RHIC energies. 
It is impressive that with a single chemical freeze-out Temperature these models can describe 
all the different particle ratios up to the heaviest strange hyperons.
Thermal parameters obtained from the fits to the various data sets seems to be extremely 
well behaved in a wide range of energy and different collisions systems and parameters 
such as Temperature and Baryo-chemical potential has been well parameterized~\cite{Cleymans0,pbm0} 
as a function of energy.
In many ways, this remarkable success of describing the data with a few simple statistical 
parameters can be viewed as an indication that indeed the particles in these collisions were 
produced in a chemically equilibrated system. 
However, if this is the case, there must be a critical system size or collision energy 
where the statistical thermal model fails to describe the observed data, where equilibration 
is not achieved due to the insufficient number of particle interactions. 
In this context, it is important to perform a careful and systematic scan of the thermal 
model fits considering systems with different size and collision energies. 
This is now possible with the large amount of data accumulated at RHIC, and in particular, 
the high quality and high statistic data obtained by the STAR experiment with its large 
acceptance detectors and extended particle identification capability.
A detailed description of the STAR detector system and the different analysis 
techniques used to obtain the yields can be found elsewhere~\cite{star0,star1,star2}. 

We present a study of the system size dependence of the hadronic freeze-out parameters by comparing 
the results from different event centrality classes of Au+Au and Cu+Cu collisions 
at $\sqrt{S_{NN}} = 200$ GeV measured by the STAR experiment.
Due to the fact that we are using data from the same detector system  many of the systematic errors 
were canceled and the relative error of the particle ratios used in the fit were reduced.
For the statistical thermal fit, we used the THERMUS code from Cleymans et al.~\cite{Cleymans1}, 
that allows for fits to the data using a Grand-Canonical ensemble approach, with the inclusion 
of a strangeness saturation parameter $\gamma_{S}$ and also a Semi-Canonical 
approach where the strangeness production was treated canonically while the light quark 
particles were still considered using a Gand-Canonical approach. 
In this study, we have used particle ratios that include the Pions, Protons, Kaons, $\Lambda$, $\Xi$, 
$\Omega$~\cite{tak1}, the corresponding anti-particles and also the $\phi$ meson~\cite{phi}.
The Pion and Proton yields were corrected for the feed-down from the weak decays before the 
thermal fit. Variations in the thermal fit parameters due to uncertainties in the feed-down 
corrections were tested and found to add an extra uncertainty of less than $5\%$ in the final 
uncertainties of these parameters.

\section{System size dependence} 
\begin{figure}
\centerline{\epsfxsize 3.1in \epsffile{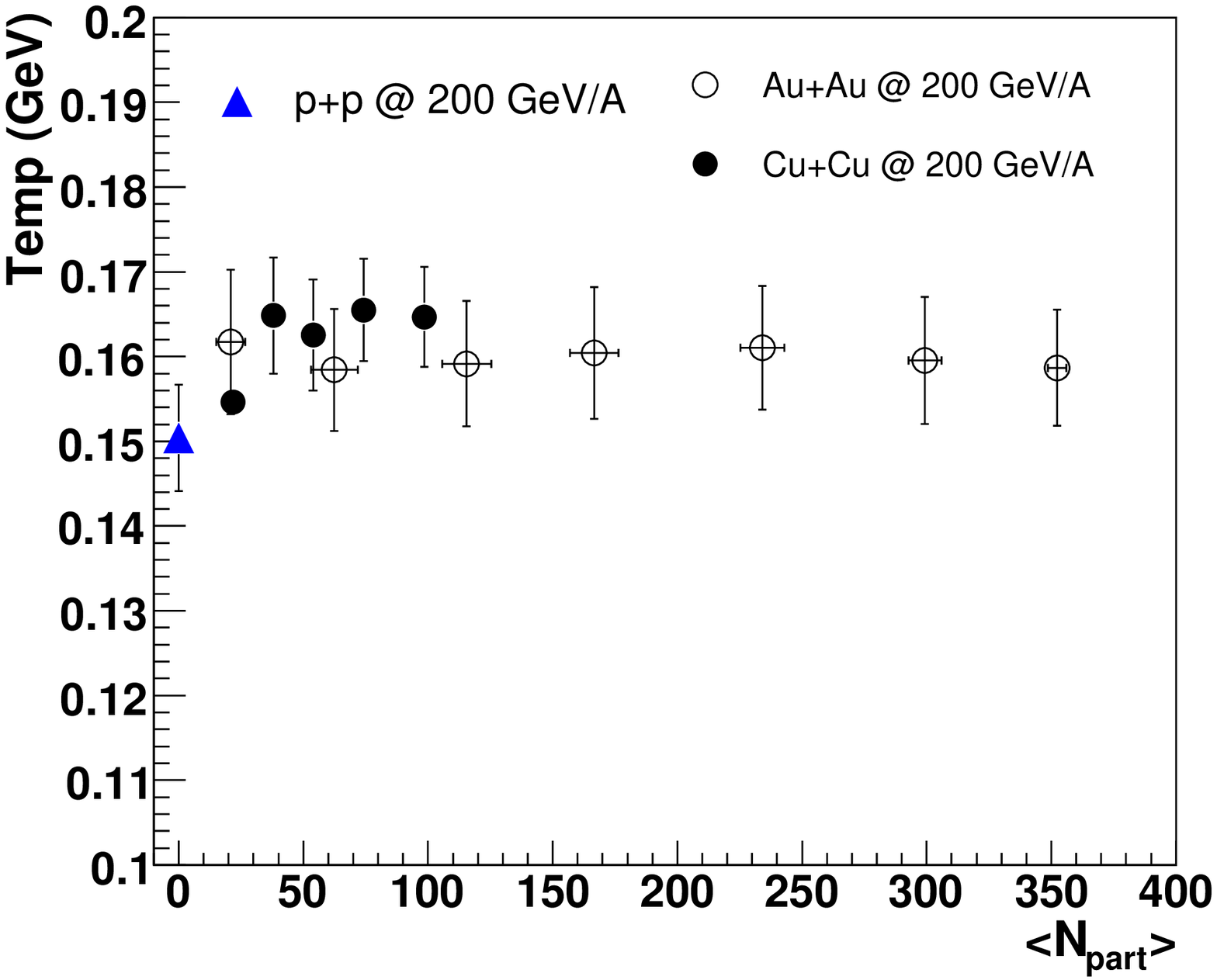} \epsfxsize 3.1in \epsffile{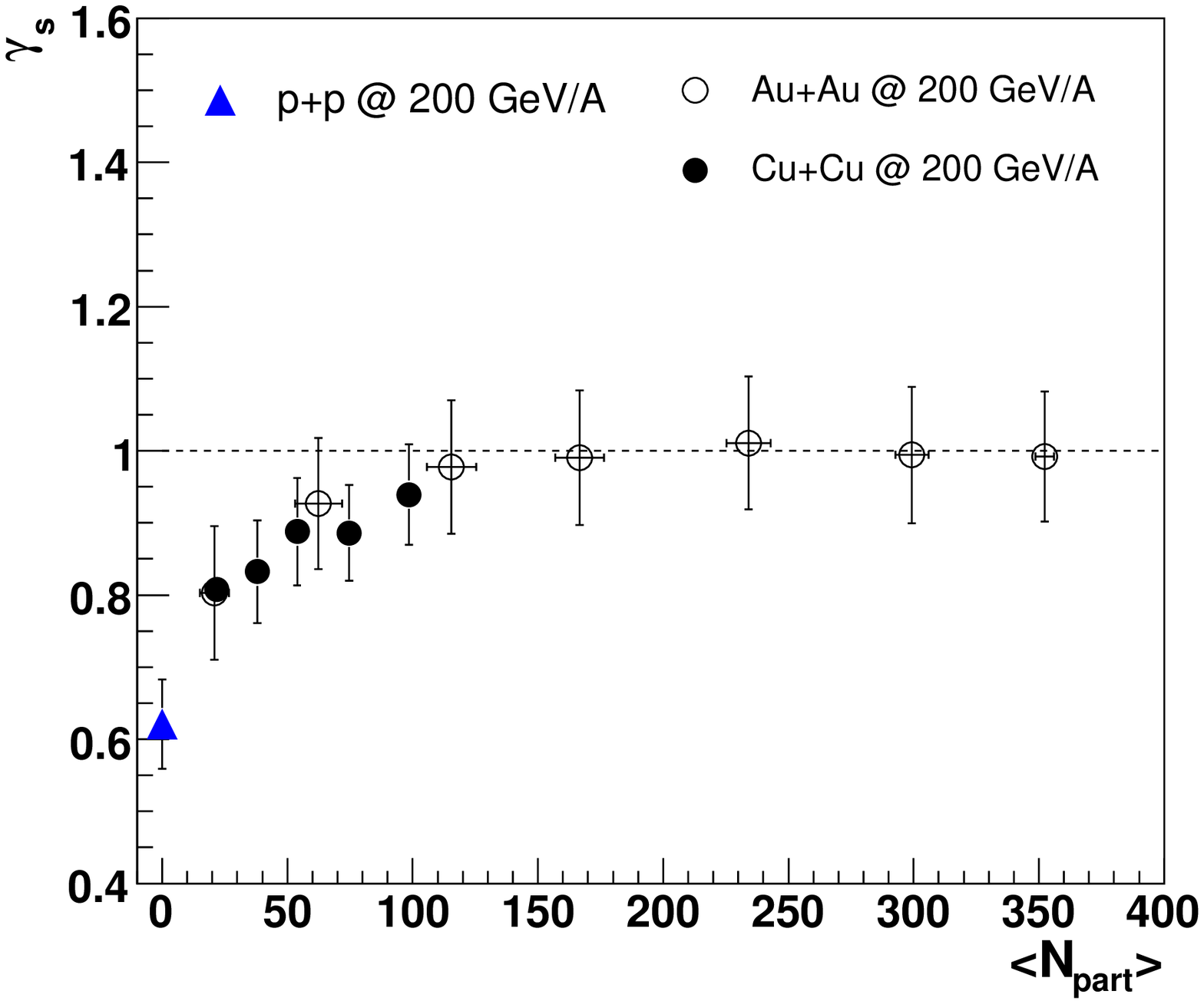}}
\caption{\label{label2} Temperature and $\gamma_{S}$ parameter as a function of the mean number of 
collisions participants $\langle N_{part}\rangle$, obtained from  p+p, Cu+Cu and Au+Au collisions at 200 GeV/A. }
\end{figure}

Particle ratios from different event centrality classes were used to study the dependence of 
chemical freeze-out conditions with the system size. Figure~\ref{label2} shows the results
of the Temperature and Strangeness saturation parameter $\gamma_{S}$ as a function of the mean 
number of participants. For comparison, data from p+p collisions was also fitted and included 
in these plots. Temperature, Baryo-chemical potential ($\mu_{B}$) and Strangeness-chemical 
potential ($\mu_{S}$) show little 
variation with the system size. Even the p+p data yields reasonable fit with the thermal model with a 
Temperature equivalent to the Au+Au data, of approximately 160 MeV. However, the $\gamma_{S}$ 
parameter shows a strong deviation from unity for smaller systems including peripheral Au+Au and Cu+Cu
events and p+p. This may be an indication that in these smaller systems, strangeness production is no 
longer well described with a simple Grand-Canonical approach. 

\begin{figure}
\centerline{\epsfxsize 3.1in \epsffile{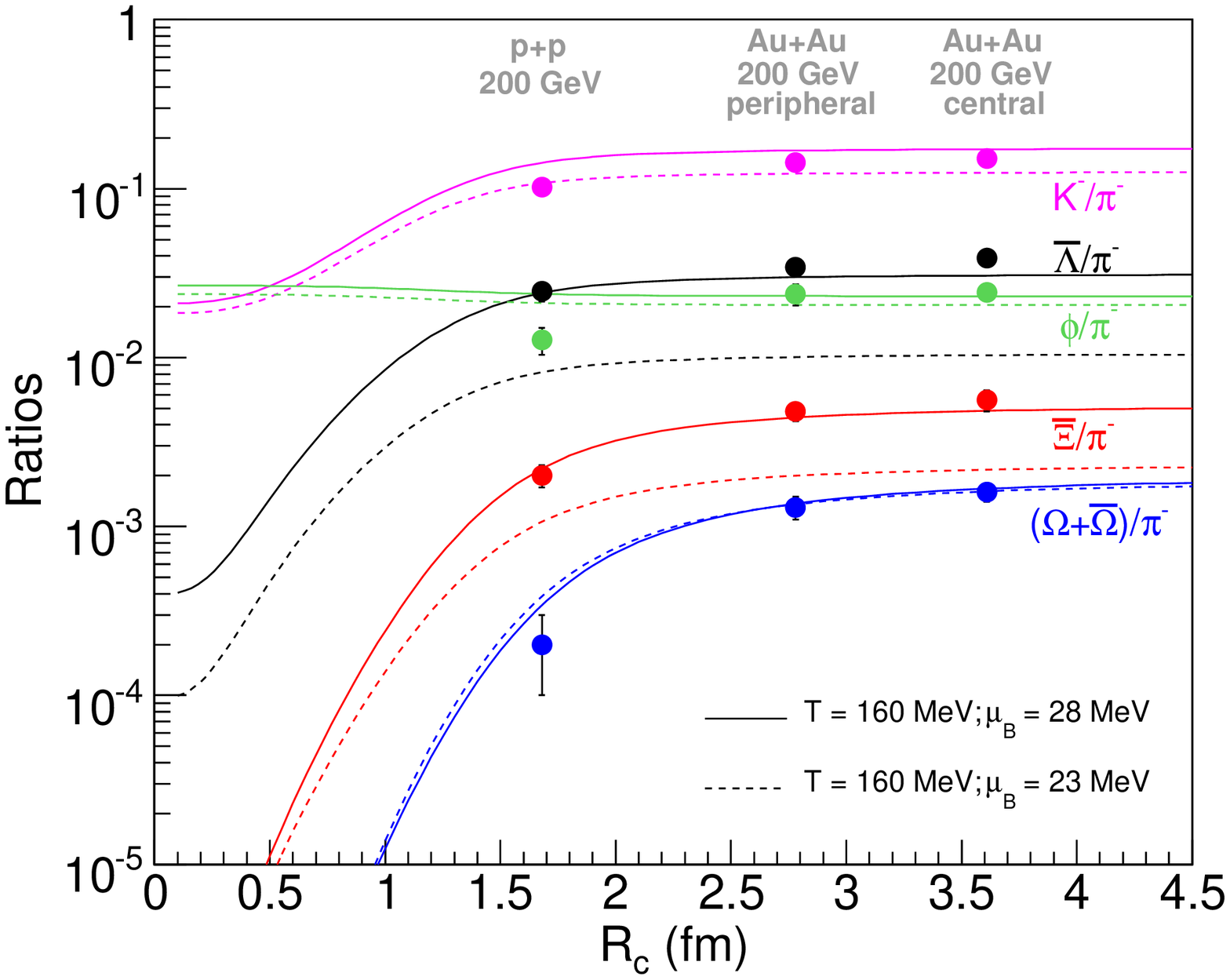} \epsfxsize 3.1in \epsffile{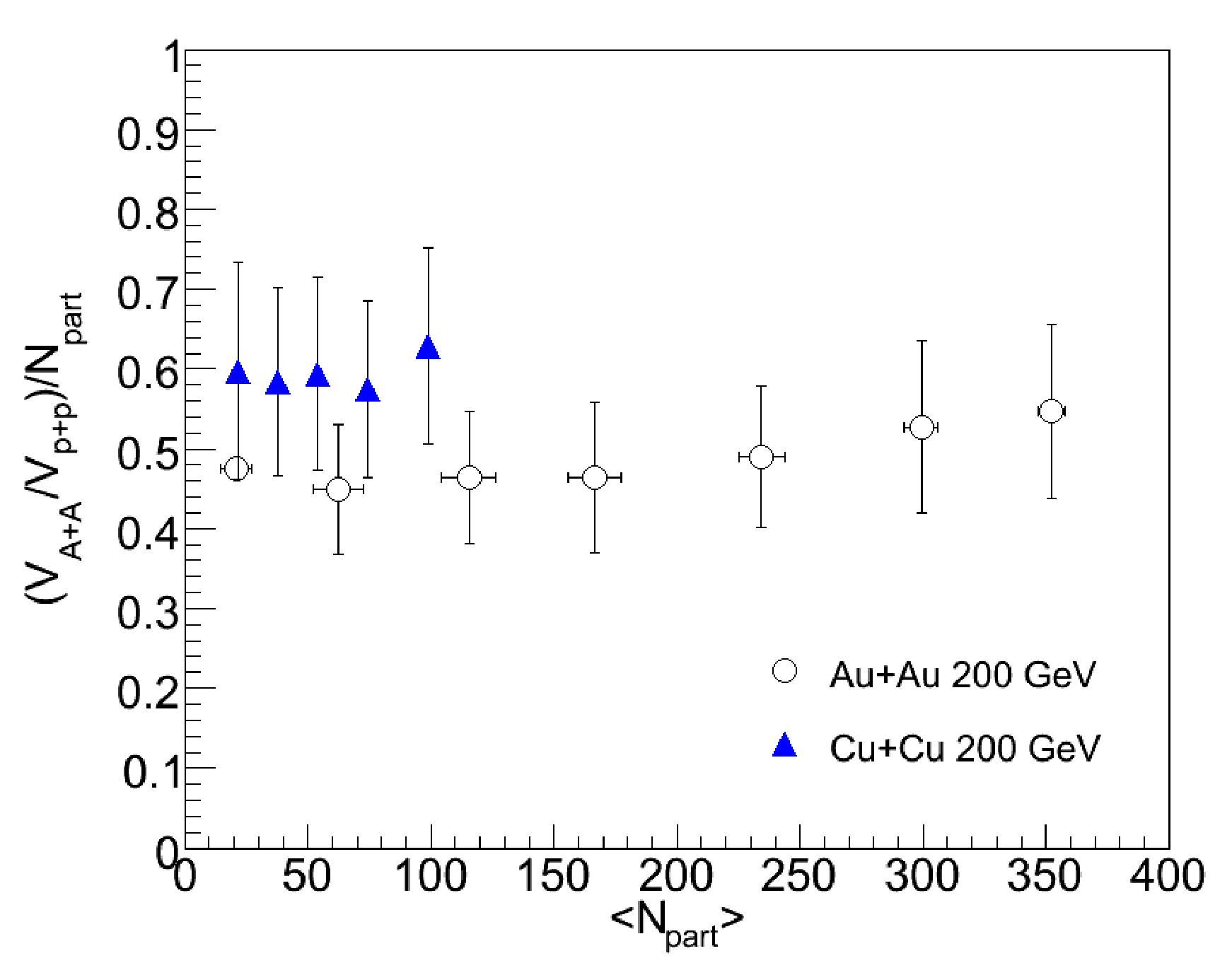} }
\caption{\label{label3}  a) Particle ratios $K^{-}/\pi$, $\bar{\Lambda}/\pi$, $\phi/\pi$, 
$\bar{\Xi}/\pi$ and $(\Omega+\bar{\Omega})/\pi$ as a function of the radius of the canonical 
correlation volume. Solid lines were obtained from the Thermal model calculation and the solid 
points represent the experimental data for central and peripheral Au+Au data and p+p data. 
b) Fraction of the system volume per participant between Au+Au and p+p extracted from the pion yields 
using the thermal fit parameters, for different event centrality classes.}
\end{figure}

To evaluate the effect of a limited system size in the strange particle production, we used a 
Semi-Canonical ensemble approach, where only the conservation of the strange sector was imposed within 
a canonical volume and the light quarks were considered still in the Grand-Canonical approach. 
Figure~\ref{label3}a shows the relative production of strange particles $K$, $\Lambda$, $\Xi$, $\Omega$ 
and the $\phi$ with respect to the pions yields, as a function of the radius of the canonical volume 
obtained using the THERMUS code. The Temperature and Baryo-chemical potential used to calculate these 
curves were fixed to be the same values obtained from the fit to the data. The solid points show the 
equivalent experimental particle ratios, for central Au+Au, peripheral Au+Au and p+p collisions.
As expected, the $\phi$ meson with its hidden strangeness shows no variation with the canonical volume, 
and the $\Omega$ shows the highest sensitivity to the canonical volume. The most peripheral data in Au+Au 
seems to yield results already consistent with values that are no longer subject to the reduction 
of strangeness production due to this canonical effect. 

The thermal fit results of the Temperature, $\mu_{B}$, $\mu_{S}$ and $\gamma_{S}$ seems to show 
no difference between the Au+Au and Cu+Cu data.  However, the relative yields per participant 
of strange particles and also of Pions, measured in Cu+Cu collisions was observed to be higher 
than the equivalent yields measured in Au+Au collisions for the same number of $\langle N_{part}\rangle$~\cite{Ant}. 
This apparent discrepancy seems not to be reflected in the particle ratios, thus, no difference in the 
thermal parameters were observed between the results from Cu+Cu and Au+Au data. 
The volume of the system at chemical freeze-out was calculated using the thermal parameters obtained from the 
fits to the particle ratios and also using the absolute yields of Pions. 
Figure~\ref{label3}b shows the relative volume per participant of the system for Au+Au and Cu+Cu data, normalized 
by the results of p+p. In this plot, we can observe that the Cu+Cu data yields a higher freeze-out volume 
than the equivalent Au+Au data.

\section{Rapidity Dependence of thermal parameters} 

Most of the results of statistical thermal fits applied to data from RHIC experiments correspond 
to particle ratios of $dN/dy$ at mid-rapidity, and not the integrated yield of particles in all 
rapidity range. To study the variation of the thermal fit parameters in the different rapidity 
range, we have used two different parameterizations of the rapidity distribution for the different 
particles to extrapolate the particle yields at the forward regions and obtain the thermal fit 
parameters. In a first attempt, we used a Gaussian shape to describe the various particle distributions, 
adjusting the yield and width of the distribution using the available data from STAR~\cite{Rafa} and 
BRAHMS~\cite{BRAHMS1, BRAHMS2} experiment. 
Particle ratios were built by extracting the corresponding $dN/dy$ for the different rapidity 
regions and also using the total integrated yield. In a second approach, we have used the well 
known HIJING Monte Carlo simulation code ~\cite{Hijing, HijingBB} to generate the particle 
rapidity distributions. The yields of the rapidity distributions generated by HIJING were scaled 
to fit the data points from STAR at mid rapidity. Once normalized, the shape of the rapidity 
distributions from HIJING seems to describe well the available data from the BRAHMS experiment 
in the forward rapidity region. 
In these fits, only particle ratios using Protons, Pions and Kaons, 
and its corresponding antiparticles were used. We noted that the main effect caused by the reduced 
number of particles in the thermal fit is the increase of the relative uncertainties in 
the final parameters and also, a reduction of the $\gamma_{S}$ parameter, from 1.0 to approximately 0.8. 

\begin{figure}
\centerline{\epsfxsize 3.1in \epsffile{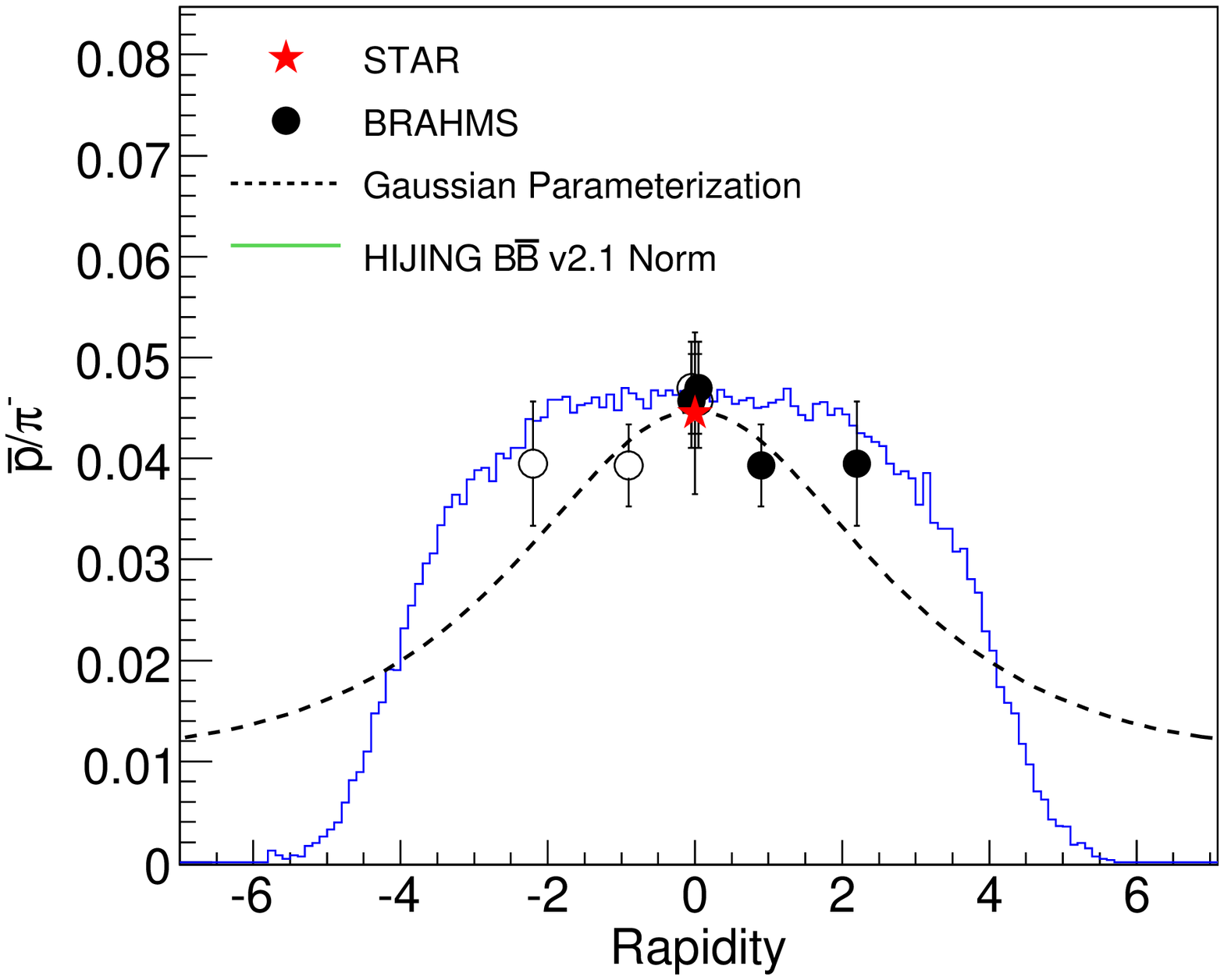} \epsfxsize 3.1in \epsffile{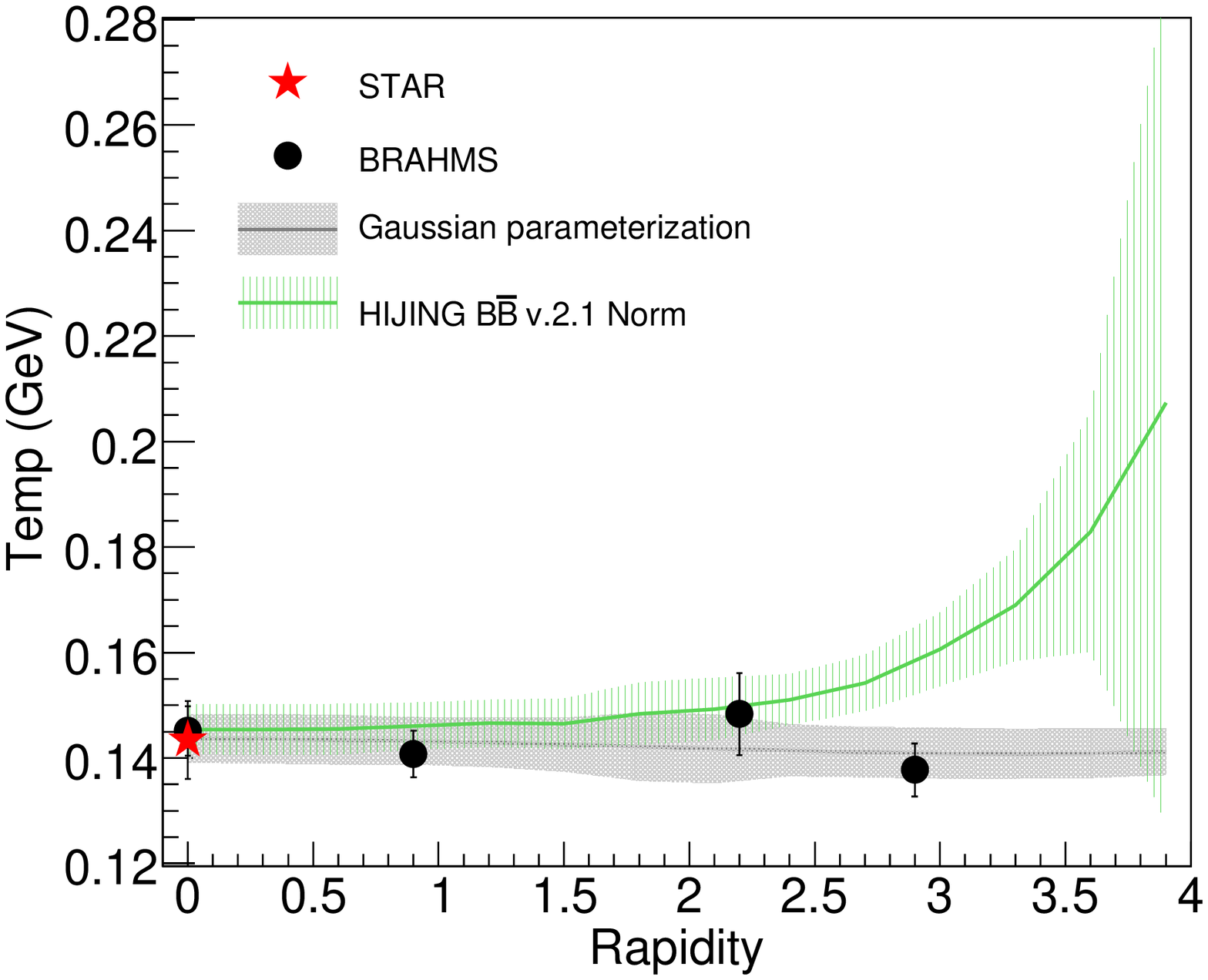}}
\caption{\label{label4} a) Particle ratio $\bar{p}/\pi$ as a function of rapidity, including 
experimental points from the STAR and BRAHMS collaboration. The dashed line represent the 
predicted ratio using a simple Gaussian distribution for describing the rapidity distribution 
of the two particles and the solid line when using HIJING.
b) Rapidity dependence of the Temperature. The points correspond to the results obtained from 
the thermal fit applied to data from STAR and BRAHMS, and the curves correspond to the results 
from fits applied to rapidity distribution obtained using a Gaussian shape or the shape from HIJING.}
\end{figure}

Figure~\ref{label4}a shows an example of the rapidity distribution obtained using the Gaussian 
extrapolation and the HIJING shape for the ratio between anti-protons and pions. In this example, 
it is clear that there are some considerable difference between the Gaussian parameterizations and 
the HIJING prediction. The solid star symbol corresponds to the STAR mid-rapidity 
measurement and the solid circles correspond to the data from the BRAHMS experiment. 
Figure~\ref{label4}b shows the Temperature obtained from the Thermal fit to this extrapolated data, and 
also the result of the fit to the STAR and BRAHMS data. Both data, and the Gaussian parameterization 
show a constant Temperature up to higher values of rapidity. The result from the thermal fit to the 
HIJING parameterization show an increase of the Temperature for the forward region with values higher 
than 2 units of rapidity.

\begin{figure}
\centerline{\epsfxsize 3.1in \epsffile{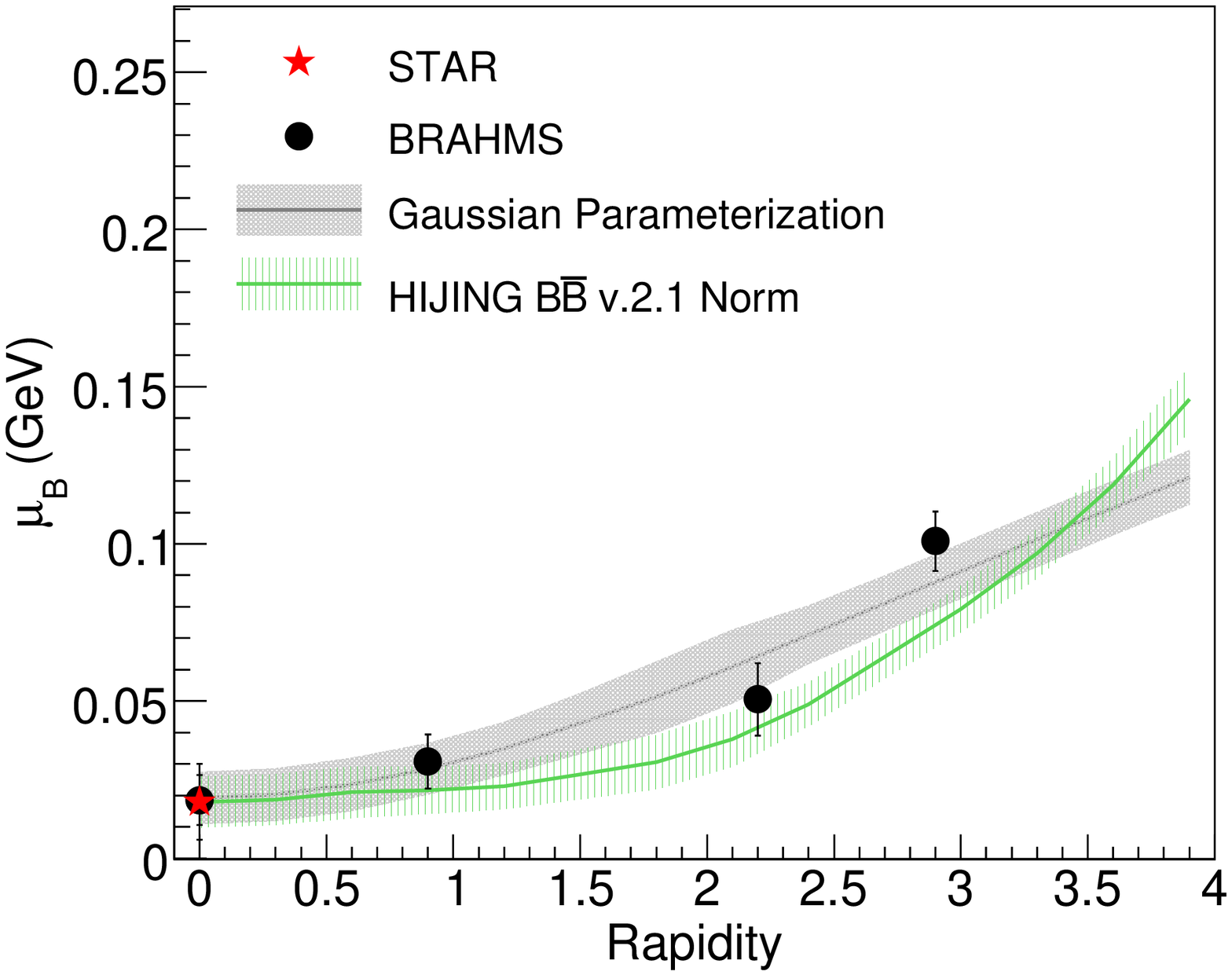} \epsfxsize 3.1in \epsffile{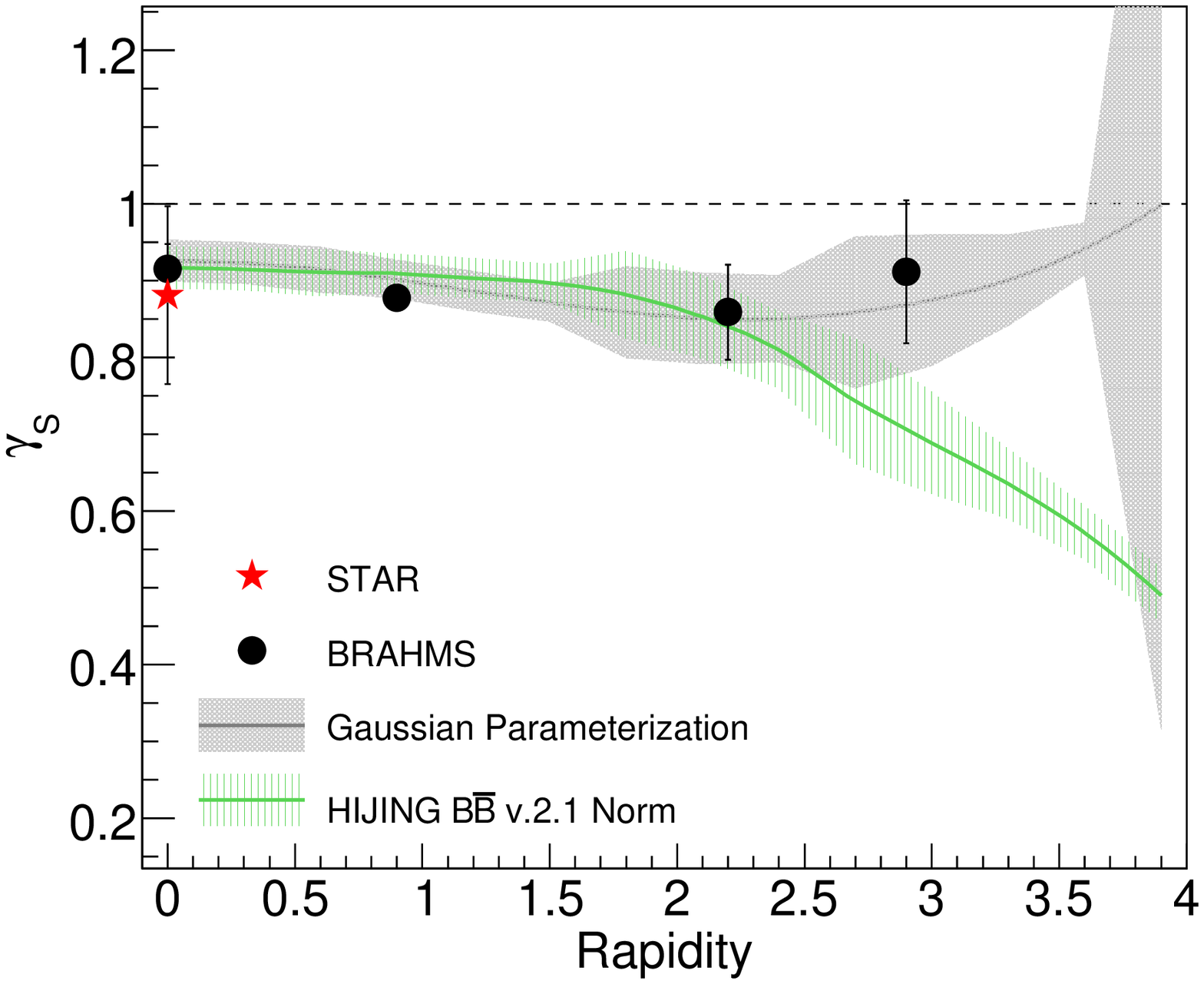}}
\caption{\label{label5} Rapidity dependence of $\mu_{B}$ and $\gamma_{s}$ from the thermal fit. 
The points correspond to the results obtained from the thermal fit applied to data from STAR and BRAHMS,
 and the curves correspond to the results from fits applied to rapidity distribution obtained using 
a Gaussian shape or the shape from HIJING.}
\end{figure}

Figure~\ref{label5}a and \ref{label5}b show the result of the Baryo-chemical potential $\mu_{B}$ and 
strangeness saturation parameter $\gamma_{S}$ as a function of rapidity. 
The Baryo-chemical potential shows an increase with rapidity reflecting the increase of the net 
baryon density of protons over antiprotons. 
The $\gamma_{S}$ parameter seems to show a constant value up to higher values of rapidity and when 
using the data and the Gaussian parameterization, but the results obtained using the HIJING extrapolations 
show a strong decrease of the strangeness equilibration in the forward direction. 
Despite the variations in the different parameterizations, all results obtained here seems to show that 
in the RHIC data, thermal characteristics of the chemical freeze-out conditions are quite constant up 
to approximately two units of rapidity. 

\section{Conclusions} 

Results from a systematic study using RHIC data for different centrality classes and different rapidity 
regions show that chemical freeze-out parameters can fit well the observed particle ratios.
The thermal parameter such as Temperature and Baryo-chemical potential show negligible variation 
with the system size, and for event classes with $N_{part} > 100$ the data is well described using 
a Gand-Canonical approach with the strangeness saturation parameter $\gamma_{S}$ consistent with unity. 
Smaller systems, such as Cu+Cu and p+p, can still be well described but with a lower value of $\gamma_{S}$. 
Using a strangeness canonical ensemble approach instead of the Grand-Canonical ensemble, we were also 
able to fit the data. The size of the canonical correlation volume for the Au+Au data is already consistent 
with a completely saturated strangeness production, where the values of the strange particle production 
no longer varies with the correlation volume. Fit to the p+p data resulted in a much smaller correlation 
volume, where canonical suppression is still a non-negligible effect into the production of strange 
particles. 
We also showed a first attempt to parameterize the rapidity dependence of the thermal fits to the 
RHIC data, where we observed that the Temperature and the $\gamma_{S}$ parameter is constant up to 
higher values of rapidity, while the Baryo-chemical potential shows an increase with rapidity.

\section*{Acknowledments} 
We wish to thank Funda\c{c}\~ao de Amparo a Pesquisa do Estado de S\~ao Paulo, FAPESP,
Brazil for the support to participate in the SQM2008 conference.

\end{document}